\documentclass[british]{article}
\usepackage[LGR,T1]{fontenc}
\usepackage[latin9]{inputenc}
\usepackage[a4paper]{geometry}
\usepackage{babel}
\usepackage{float}
\usepackage{mathtools}
\usepackage{url}
\usepackage{dsfont}
\usepackage{amsmath}
\usepackage{amssymb}
\usepackage{graphicx}
\usepackage{setspace}
\usepackage[numbers]{natbib}
\usepackage[bookmarks=true,bookmarksnumbered=false,bookmarksopen=false,
 breaklinks=false,pdfborder={0 0 1},backref=false,colorlinks=false]
 {hyperref}
\hypersetup{
 colorlinks=true,citecolor=magenta,linkcolor=blue}

\makeatletter

\DeclareRobustCommand{\greektext}{%
  \fontencoding{LGR}\selectfont\def\encodingdefault{LGR}}
\DeclareRobustCommand{\textgreek}[1]{\leavevmode{\greektext #1}}

\newcommand{\lyxaddress}[1]{
	\par {\raggedright #1
	\vspace{1.4em}
	\noindent\par}
}

\@ifundefined{date}{}{\date{}}

\usepackage{authblk}

\makeatother

\begin{document}
\title{Optimising Recruitment and Sample Allocation in Multi-Site Rare Disease
Studies}
\author{Farid Zare\textsuperscript{1, 2, 3}, Catherine Clancy\textsuperscript{1, 2, 3},
Tal Dattner\textsuperscript{4}, Martina Messina\textsuperscript{5, 6},
Julian Teinert\textsuperscript{4}, Jack McGoldrick\textsuperscript{1, 2, 3},
Pavel Jesina\textsuperscript{7}, Sabine Weller Gr\o{}nborg\textsuperscript{8},
Ute Spiekerkoetter\textsuperscript{9}, James J O'Byrne\textsuperscript{10},
Trine Tangeraas\textsuperscript{11, 12}, Maria Luz Couce\textsuperscript{13},
Jos\'e A. Cocho\textsuperscript{13}, Javier de las Heras\textsuperscript{14},
Niklas Darin\textsuperscript{15}, Dorothea M\"o{}slinger\textsuperscript{16},
Peter Witters\textsuperscript{17}, Ina Knerr\textsuperscript{18},
Giancarlo La Marca\textsuperscript{19, 20}, Ans T. van der Ploeg\textsuperscript{21},
Hidde H. Huidekoper\textsuperscript{21}, Terry G. J. Derks\textsuperscript{22},
Francjan J. van Spronsen\textsuperscript{12, 22}, Sandra D.K Kingma\textsuperscript{23},
Anibh M. Das\textsuperscript{24}, Carlo Dionisi-Vici\textsuperscript{25},
Clara D. van Karnebeek\textsuperscript{26}, Julia B. Hennermann\textsuperscript{27},
Siobh\'a{}n O'Sullivan\textsuperscript{28}, Sabine A. Fuchs\textsuperscript{29},
Luciana Hannibal\textsuperscript{30, 31, 32}, M. Rebecca Heiner-Fokkema\textsuperscript{33},
Marek J. Noga\textsuperscript{34}, Irene M. L. W. K\"o{}rver-Keularts\textsuperscript{34},
Katja Benedikte Prest\o{} Elgst\o{}en\textsuperscript{35, 36},
George J.G. Ruijter\textsuperscript{37}, J J M Jans\textsuperscript{38},
Amy Harms\textsuperscript{39}, Thomas Hankemier\textsuperscript{39},
Maurizio Scarpa\textsuperscript{40}, Ines Thiele\textsuperscript{1, 2, 3},
Shamima Rahman\textsuperscript{5, 6}, Stefan K\"o{}lker\textsuperscript{4},
Ronan M.T. Fleming\textsuperscript{1, 2, 3 }}
\maketitle
\begin{singlespace}

\lyxaddress{\textsuperscript{1}School of Medicine, University of Galway, University
Rd, Galway, Ireland; \textsuperscript{2}Institute for Clinical Trials,
University of Galway, Galway, Ireland; \textsuperscript{3}Digital
Metabolic Twin Centre, University of Galway, Galway, Ireland; \textsuperscript{4}Heidelberg
University, Medical Faculty, and University Hospital Heidelberg, Centre
for Paediatrics and Adolescent Medicine, Division of Paediatric Neurology
and Metabolic Medicine, Heidelberg University, Heidelberg, Germany;
\textsuperscript{5}Mitochondrial Research Group, Genetics and Genomic
Medicine Department, UCL Great Ormond Street Institute of Child Health,
London, UK; \textsuperscript{6}Metabolic Unit, Great Ormond Street
Hospital for Children NHS Foundation Trust, London, UK; \textsuperscript{7}Department
of Pediatrics and Inherited Metabolic Disorders, First Faculty of
Medicine, Charles University and General University Hospital, Prague,
Czech Republic; \textsuperscript{8}Centre for Inherited Metabolic
Diseases, Department of Paediatrics and Adolescent Medicine and Department
of Clinical Genetics, Copenhagen University Hospital, Rigshospitalet,
Copenhagen, Denmark; \textsuperscript{9}Department of General Paediatrics,
Adolescent Medicine and Neonatology, Faculty of Medicine, University
Medical Centre, University of Freiburg, Freiburg, Germany; \textsuperscript{10}National
Centre for Inherited Metabolic Disorders, Mater Misericordiae University
Hospital, Dublin, Ireland; \textsuperscript{11}Department of Paediatrics
and Adolescent Medicine, Oslo University Hospital, Oslo, 0424, Norway;
\textsuperscript{12}European Reference Network for Hereditary Metabolic
Disorders (MetabERN), Udine, Italy; \textsuperscript{13}Unit of Metabolic
Diseases, University Clinical Hospital of Santiago, MetabERN, IDIS,
15706 Santiago de Compostela, Spain; \textsuperscript{14}Cruces University
Hospital, CIBERER, University of the Basque Country (EHU) and Biobizkaia
Health Research Institute ES; \textsuperscript{15}Department of Paediatrics,
Institute of Clinical Sciences, Sahlgrenska Academy, University of
Gothenburg, 40530 Gothenburg, Sweden; \textsuperscript{16}Department
of Paediatrics and Adolescent Medicine, Medical University Vienna,
Vienna, Austria; \textsuperscript{17}Center for Metabolic Diseases,
Department of Paediatrics, University Hospitals Leuven, Leuven, Belgium;
\textsuperscript{18}National Centre for Inherited Metabolic Disorders,
Children\textquoteright s Health Ireland at Temple Street, Dublin,
Ireland; \textsuperscript{19}Newborn Screening, Clinical Biochemistry
and Clinical Pharmacy Lab, Meyer Children\textquoteright s Hospital
IRCCS, Florence, Italy; \textsuperscript{20}Department of Experimental
and Clinical Biomedical Sciences, University of Florence, Florence,
Italy; \textsuperscript{21}Erasmus MC, University Medical Centre
Rotterdam, Rotterdam, The Netherlands; \textsuperscript{22}Section
of Metabolic Diseases, Beatrix Children\textquoteright s Hospital,
University Medical Centre Groningen, University of Groningen, Groningen,
The Netherlands; \textsuperscript{23}Center of Inherited Metabolic
Diseases, Antwerp University Hospital, University of Antwerp, Edegem,
Belgium; \textsuperscript{24}Clinic for Paediatric Kidney, Liver
and Metabolic Diseases and Neuropaediatrics, Hannover Medical School,
Hannover, Germany; \textsuperscript{25}Division of Metabolic Diseases
and Hepatology, Bambino Ges\`u{} Children\textquoteright s Hospital
IRCCS, Rome, Italy; \textsuperscript{26}Department of Paediatrics,
Amsterdam Gastroenterology Endocrinology Metabolism, Emma Children\textquoteright s
Hospital, Amsterdam UMC; Amsterdam Reproduction \& Development Research
Institute, Child Development, Amsterdam, The Netherlands; \textsuperscript{27}University
Medical Centre Mainz, Centre for Paediatric and Adolescent Medicine,
Villa Metabolica, Mainz, Germany; \textsuperscript{28}Department
of Paediatrics, Royal Belfast Hospital for Sick Children, Belfast,
UK; \textsuperscript{29}Department of Metabolic Diseases, Wilhelmina
Children\textquoteright s Hospital, University Medical Centre Utrecht,
Utrecht, The Netherlands; \textsuperscript{30}Laboratory of Clinical
Biochemistry and Metabolism, Department of General Pediatrics, Adolescent
Medicine and Neonatology, Faculty of Medicine, Medical Center, University
of Freiburg, 79106 Freiburg, Germany; \textsuperscript{31}Translational
Metabolomics Core Facility, Faculty of Medicine, Medical Center --
University of Freiburg, 79106 Freiburg, Germany; \textsuperscript{32}CIBSS
-- Centre for Integrative Biological Signalling Studies, University
of Freiburg, 79106 Freiburg, Germany; \textsuperscript{33}Laboratory
of Metabolic Diseases, Department of Laboratory Medicine, University
of Groningen, University Medical Centre Groningen, 9700 RB Groningen,
The Netherlands; \textsuperscript{34}Laboratory of Clinical Genetics,
Maastricht University Medical Center, Maastricht, The Netherlands;
\textsuperscript{35}Department of Medical Biochemistry, Oslo University
Hospital -- Rikshospitalet, Oslo, Norway; \textsuperscript{36}Core
Facility for Global Metabolomics and Lipidomics, Faculty of Medicine,
University of Oslo, Oslo, Norway; \textsuperscript{37}Department
of Clinical Genetics, Erasmus MC University Medical Centre, Rotterdam,
The Netherlands; \textsuperscript{38}Department of Genetics, Section
Metabolic Diagnostics, University Medical Centre Utrecht, Utrecht,
The Netherlands; \textsuperscript{39}Metabolomics and Analytics Centre,
Leiden Academic Centre for Drug Research (LACDR), Leiden University,
The Netherlands; \textsuperscript{40}MetabERN, Regional Coordinating
Centre for Rare Diseases, University Hospital, Udine, Italy.}
\end{singlespace}

\lyxaddress{Correspondence: R.M.T. Fleming, ronan.mt.fleming@gmail.com}
\begin{abstract}
Clinical research in rare diseases is challenged by the limited availability
and broad geographical distribution of eligible patients. This necessitates
multi-centre collaboration and strategic selection of analytical laboratories
based on capacity, cost, and capability. We present a mathematical
framework based on cardinality optimisation to support key operational
decisions in study design, balancing clinical relevance, feasibility,
budgetary constraints, and logistical factors. The framework was implemented
in the Reconstruction and Computational Modelling for Inherited Metabolic
Diseases (Recon4IMD) project, aiming to recruit 1,900 diagnosed patients
across 24 healthcare centres and eight metabolomic laboratories. Compared
to both proportional heuristic and greedy marginal allocation baselines,
the proposed optimisation model consistently achieved higher recruitment
yields in simulated multi-centre scenarios and reduced metabolomic
analysis costs by approximately 30\% under unconstrained conditions.
Notably, both the heuristic and greedy allocation baselines became
infeasible under progressively tighter laboratory capacity constraints,
whereas the optimised model maintained feasibility and cost control.
These findings highlight the value of integrated optimisation strategies
for improving recruitment, platform selection, and sample allocation
in complex, resource-limited research settings.
\end{abstract}
keywords: Cardinality optimisation, Rare disease, Patient recruitment
strategy, Decision-making, Large-scale optimisation

\section*{Introduction}

Inherited metabolic disorders (IMDs) are rare genetic diseases that
pose significant recruitment challenges due to their geographic dispersion
and the need for specialised laboratories for diagnosis and monitoring.
Multi-centre clinical studies have become indispensable in medical
research, enabling faster recruitment and more generalisable results
than single-site studies \citep{harden_sample_2018}. In rare diseases,
no single hospital can enrol a sufficiently large cohort for meaningful
results, making collaboration across multiple sites essential for
achieving adequate sample sizes \citep{zurek_solve-rd_2021}. However,
these advantages come with operational challenges, particularly in
patient recruitment and sample allocation.

Recruitment in rare disease studies often remains slow, with each
site typically contributing only a few patients. Ensuring consistent
patient selection and diagnostic criteria across sites adds complexity,
especially when expert consensus is required. Designing a multi-centre
study thus involves balancing the inclusion of many centres against
slow, diffuse recruitment \citep{hilgers_design_2016}. Laboratory
selection also matters: inter-site assay variability is common \citep{lee_harmonization_2018},
prompting many studies to centralise testing or harmonise protocols
to improve data quality. Previous research has produced piecemeal
solutions for these issues, including better sample size formulas
and calls for centralised laboratories \citep{burmeister_sample_2012}.

Multi-site studies are resource-intensive, with regulatory, ethical,
and logistical requirements often limiting expansion \citep{tang_challenges_2019}.
Continuous management of recruitment strategies and data across sites
is needed to keep studies on track \citep{myers_recruitment_2019}.
Together, these challenges make multi-centre study planning a complex,
multi-factorial decision problem.

Mathematical optimisation offers a powerful framework for addressing
such complexity, formulating problems as the minimisation or maximisation
of objectives under constraints. Within this, cardinality optimisation
focuses on selecting an optimal subset of elements, minimising or
maximising the number of sites, labs, or allocations. This approach,
which enforces parsimony in solutions, has been effective in computational
biology, such as in metabolic modelling \citep{fleming_cardinality_2023a},
and shows promise for high-dimensional decision spaces.

Despite these advances, cardinality optimisation has not yet been
integrated into operational decision-making for clinical research.
While computational frameworks exist for simulating studies \citep{kozlowska_mathematical_2024},
there is a lack of applications addressing practical multi-centre
questions, such as site selection, lab allocation, or logistical minimisation
in a unified way. Clinical study planning still relies on heuristics,
isolated analyses, and scenario iteration, rather than a comprehensive
algorithmic framework. This gap presents an opportunity to improve
multi-centre study design by integrating cardinality optimisation.

Traditional planning approaches often decompose study design into
sequential decisions, such as defining sample-size targets, estimating
recruitment capacity, and then addressing operational or logistical
requirements separately \citep{harden_sample_2018,tang_challenges_2019,myers_recruitment_2019}.
This separation is practical but limited when decisions are interdependent.
In multi-site rare disease studies, recruitment at one healthcare
provider changes disease-specific cohort totals, downstream biomarker
demand, analytical platform use, laboratory capacity, cost, and sample
shipment patterns across the network. Laboratory allocation problems
also introduce cost and capacity constraints that are difficult to
address through recruitment planning alone \citep{davila_laboratory_2018}.
Greedy strategies tend to prioritise local availability or immediate
marginal gain, without jointly accounting for cross-disease dependencies,
shared capacity constraints, clinical interest, and logistical trade-offs
\citep{chvatal_greedy_1979,nemhauser_analysis_1978}. These limitations
motivate an integrated optimisation framework in which recruitment,
biomarker demand, platform allocation, cost, and logistics are represented
within a single constrained model.

The optimisation framework integrates recruitment planning and sample
allocation within a single mathematical model. Disease-specific patient
availability across healthcare providers is combined with recruitment
capacity, cohort-size targets, biomarker requirements, analytical
platform capabilities, laboratory throughput, analysis costs, logistical
constraints, and hospital-specific clinical interests. Recruitment
decisions define disease-specific cohort sizes and determine biomarker
measurement demand, which is then linked to allocation across targeted
and global metabolomic platforms. The unified objective combines cardinality
terms, recruitment yield, logistical deviation, cost, and clinical-interest
alignment to identify feasible and efficient multi-site recruitment
and sample allocation strategies. Using the Reconstruction and Computational
Modelling for Inherited Metabolic Diseases (Recon4IMD) project as
a case study, we show how this strategy supports cost-efficient, logistically
feasible, and clinically robust decision-making in complex multi-centre
studies. The full mathematical formulation is provided in the Methods
section.
\section*{Results}

The following sections report the application of the proposed mathematical
modelling framework to the Recon4IMD project. We first identify which
IMDs are feasible for inclusion, based on combined genetic, metabolomic,
and recruitment criteria. We then present an overview of patient recruitment
and sample allocation across participating healthcare providers and
metabolomic platforms. Performance benchmarking compares the cardinality-optimised
approach to proportional heuristic and greedy baseline strategies
for both recruitment and sample allocation, evaluating recruitment
yield, cost efficiency, and feasibility under varying laboratory constraints.
Finally, we analyse trade-offs between clinical alignment and logistics,
highlighting the flexibility of the optimisation model in balancing
competing objectives.

\subsection*{Feasible inherited metabolic diseases in the Recon4IMD project}

\begin{figure}[H]
\centering
\includegraphics[scale=0.45]{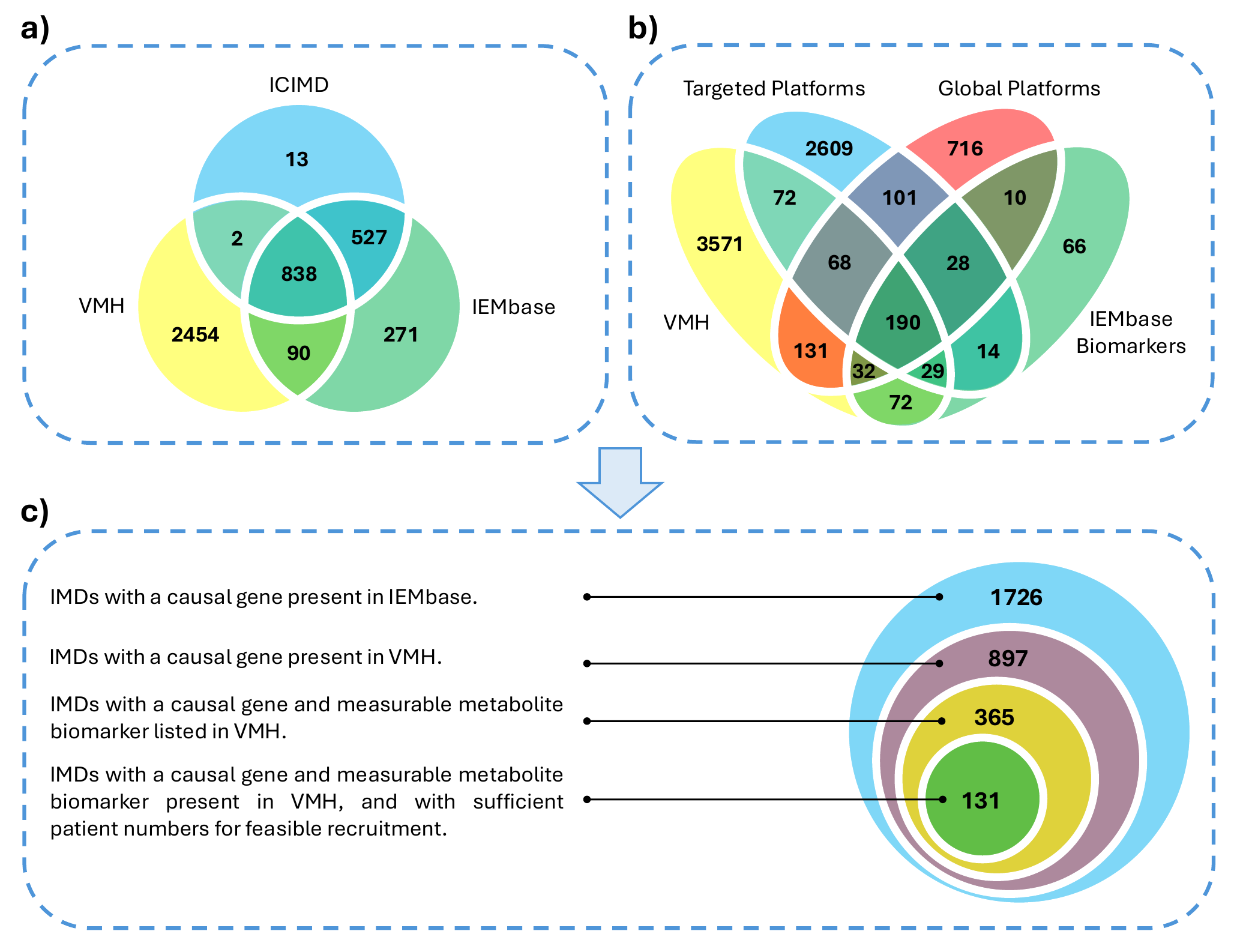}

\caption{Integration and filtering of inherited metabolic disease data across
databases and platforms. This figure illustrates the harmonisation
of genetic and metabolic data from multiple sources to identify suitability
for metabolic modelling. a) Overlap of gene sets among the VMH, ICIMD,
and IEMbase databases using gene symbols as the common namespace.
b) Overlap of metabolite sets among the VMH, IEMbase metabolite biomarkers,
and targeted and global metabolomic platforms. Comparison is performed
using MetaNetX identifiers. c) Stepwise filtering of Inborn Errors
of Metabolism (IMDs) from the IEMbase database. The filtering process
evaluates data suitability for metabolic modelling and feasibility
for patient recruitment based on gene and biomarker availability in
VMH and the presence of measurable metabolite biomarkers.\protect\label{fig:Stepwise-filtering-of-IMDs}}
\end{figure}

Three databases, VMH, IEMbase, and ICIMD, were investigated and compared
for inclusion in Recon4IMD. The comparison focuses on the overlap
of causal IMD genes and known metabolite biomarkers across the datasets.
Figure \ref{fig:Stepwise-filtering-of-IMDs}a shows a Venn diagram
illustrating the overlap of gene sets across the three databases,
using gene symbols as the common namespace. The VMH database contains
approximately 60\% of the IMD causal genes listed in ICIMD, and over
50\% of those catalogued in IEMbase. Notably, IEMbase includes over
99\% of the causal genes present in ICIMD. All three databases share
a core set of 838 genes.

Similarly, Figure \ref{fig:Stepwise-filtering-of-IMDs}b compares
metabolite sets, including those in VMH, metabolites measurable via
targeted and global platforms, and IMD-specific biomarker metabolites
listed in IEMbase. A subset of 190 intersecting metabolites represents
those that are included in the genome-scale models, are biomarkers
for IMDs, and are measurable through existing targeted and global
metabolomic platforms within the Recon4IMD project.

By requiring that an IMD has its causal gene and at least one metabolite
biomarker represented in the VMH database, and that its biomarker
metabolites are measurable using available analytical platforms, we
identified the set of IMDs feasible for inclusion in the Recon4IMD
project. Figure \ref{fig:Stepwise-filtering-of-IMDs}c illustrates
this stepwise filtering process. The outermost circle (n = 1726) includes
all IMDs in IEMbase that are annotated with at least one causal gene.
The next subset (n = 897) includes IMDs with at least one causal gene
present in the VMH database. The third circle (n = 365) further restricts
the set to IMDs with at least one measurable metabolite biomarker.

In total, 365 IMDs were identified as suitable for modelling based
on the presence of at least one causal gene in the VMH database and
at least one measurable metabolite biomarker. However, recruitment
feasibility was an additional constraint: not all of these IMDs had
sufficient numbers of diagnosed patients across the 24 participating
healthcare providers. By incorporating recruitment data and clinical
insights, a final subset of 131 IMDs was selected, representing diseases
with modellable genes and metabolites, with sufficient patient numbers
to enable feasible recruitment within the context of the Recon4IMD
project, while excluding entities that were merely biochemical traits
without clinical correlates (i.e. 'non-diseases'). These results reflect
the database contents at the time the work was conducted and may evolve
as VMH, IEMbase, and ICIMD are updated.

\subsection*{Recruitment and participation rate}

\begin{figure}[H]
\centering
\includegraphics[scale=0.6]{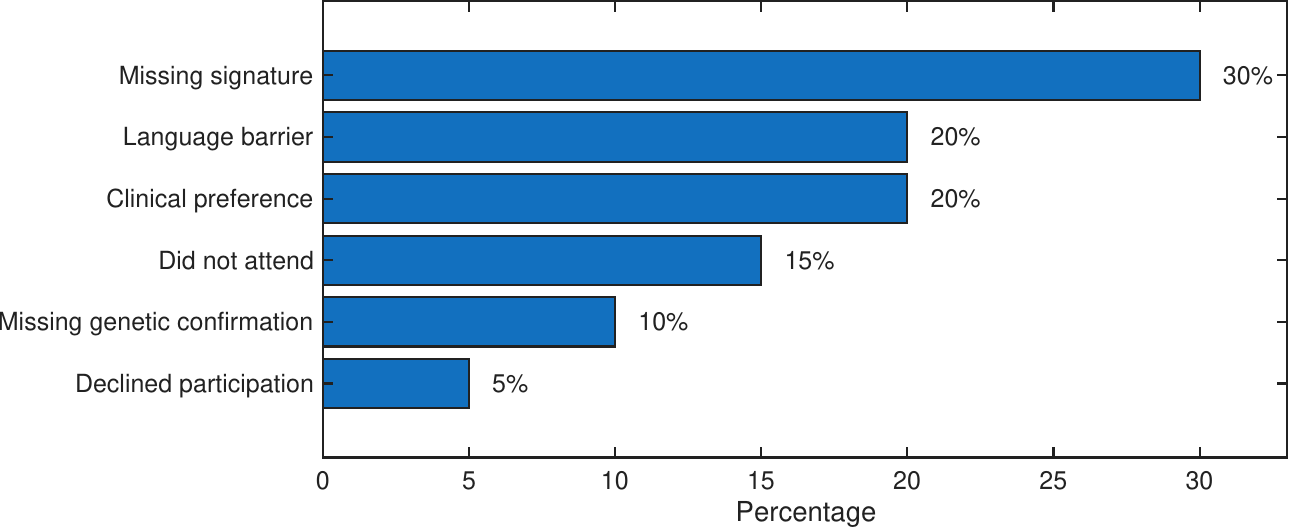}\caption{Reasons for non-inclusion in the study. Bars show the proportion of
assessed participants who were not enrolled, ordered from most to
least frequent: one-parent consent, language barrier, too unwell or
no planned blood draw, did not attend, no genetic test in legacy PA/MSUD
cases, and declined participation.\protect\label{fig:Reasons-for-non-inclusion}}
\end{figure}

During the first four months of recruitment at Heidelberg University
Hospital, we recorded patient consent outcomes. We also documented
reasons for failed sample collection. Of 234 patients approached,
113 consented to provide blood and/or urine samples, giving a participation
rate of 48.3\%, which was close to the initial assumption of 50\%.
Figure \ref{fig:Reasons-for-non-inclusion} shows the reasons for
non-inclusion. Among the 121 non-included patients, the most frequent
reasons were missing signature (30\%), reflecting the legal requirement
in Germany for both parents to provide consent for minors; language
barrier (20\%), where informed consent could not be obtained due to
communication difficulties; and clinical preference (20\%), when treating
physicians decided not to approach an eligible patient in order to
avoid overburdening or to minimise blood draws in children. These
top three reasons accounted for 70\% of non-inclusions. The remaining
cases were attributed to did not attend (15\%), when patients failed
to appear for scheduled visits; missing genetic confirmation (10\%),
referring to patients without confirmed molecular diagnosis such as
legacy PA/MSUD cases; and declined participation (5\%), when the patient
or family actively refused. Grouped by theme, structural or regulatory
factors represented 40\% (missing signature, missing genetic confirmation),
clinical factors 20\% (clinical preference), logistical or communication
barriers 35\% (language barrier, did not attend), and patient preference
5\% (declined). Notably, the requirement for consent from both parents
applies specifically to paediatric participants in Germany and may
differ in other countries.

\subsection*{Visualising IMD coverage and sample allocation across the network}

\begin{figure}[H]
\centering
\includegraphics[scale=0.051]{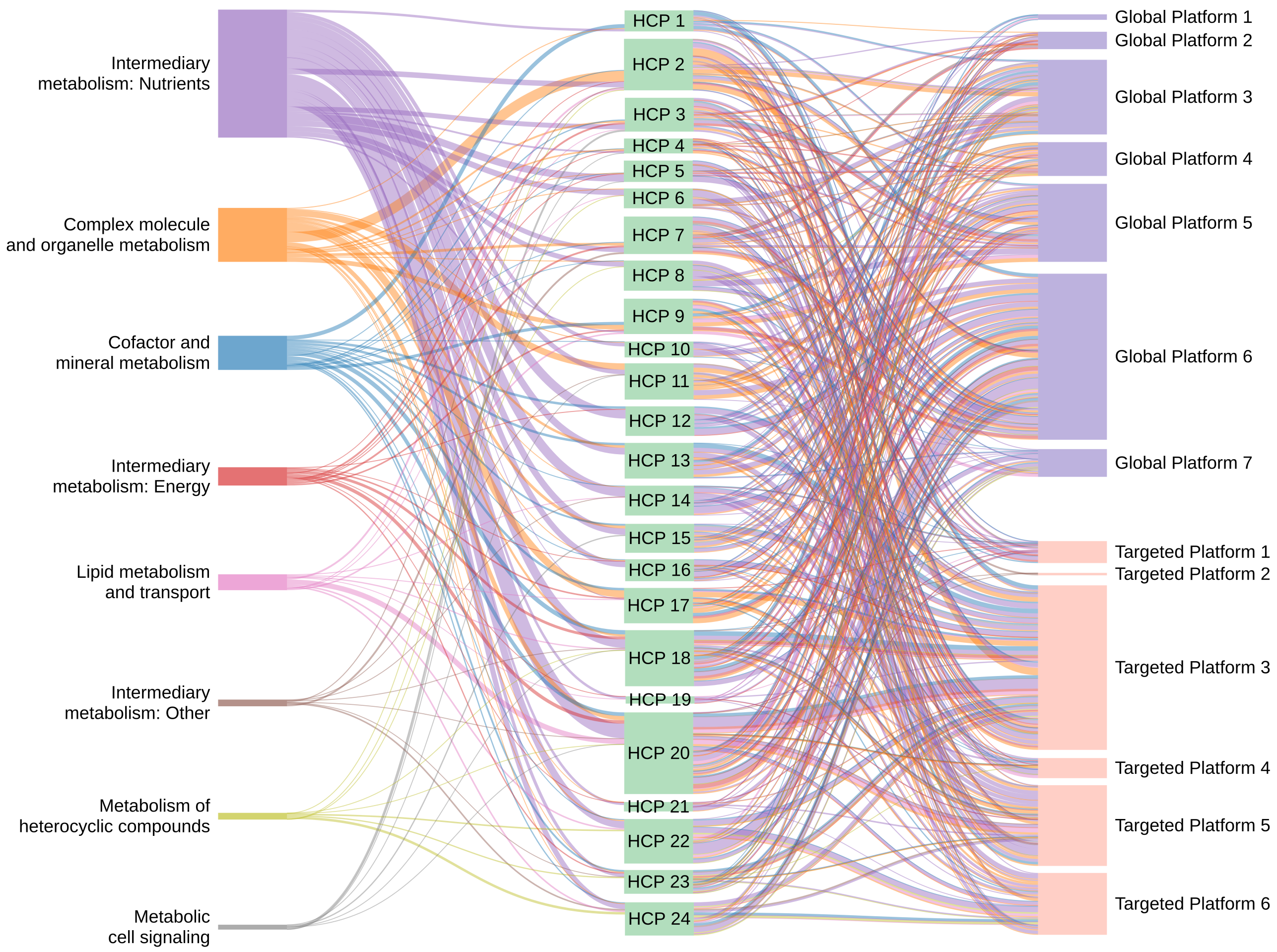}

\caption{Optimised sample allocation from ICIMD disease groups to healthcare
providers (HCPs) and analytical platforms in the Recon4IMD project.
This Sankey diagram visualises the flow of patient samples from inherited
metabolic disorders, categorised by ICIMD disease groups (left), through
24 healthcare providers (centre), to analytical platforms (right).
Each flow represents the number of patients recruited and assigned
for omics analysis under the cardinality-optimised model. Samples
are directed to either global platforms (purple) or targeted platforms
(blue), based on biomarker coverage, laboratory capacity, and logistical
constraints. Flow colours correspond to ICIMD groups, highlighting
how disease categories are distributed across sites and platforms.
The diagram illustrates the integrated recruitment and sample allocation
strategy designed to maximise diagnostic yield while minimising logistical
burden.\protect\label{fig:sankey-diagram}}
\end{figure}

To visualise how patient recruitment and sample allocation unfold
across disease categories, sites, and platforms under the optimised
strategy, we constructed a Sankey diagram. Figure \ref{fig:sankey-diagram}
reveals the interplay between ICIMD disease groups, the 24 healthcare
providers (HCPs), and the designated analytical platforms. It highlights
how disease complexity is managed through coordinated platform use,
while balancing biomarker coverage, laboratory constraints, and logistics.

\subsection*{Benchmarking results}

To benchmark recruitment performance, we simulated 24 scenarios with
increasing numbers of HCPs, ranging from single-centre to full-network
participation. In each scenario, the cardinality-optimised model was
compared against two baseline recruitment strategies: the proportional
heuristic approach (Algorithm S1) and the greedy recruitment baseline
(Algorithm S2). All methods were applied to the same set of hospitals
in each scenario, and the total number of recruited patients was recorded.
This comparison focused exclusively on recruitment performance and
did not incorporate downstream allocation or cost considerations.

\begin{figure}[H]
\centering
\includegraphics[scale=0.6]{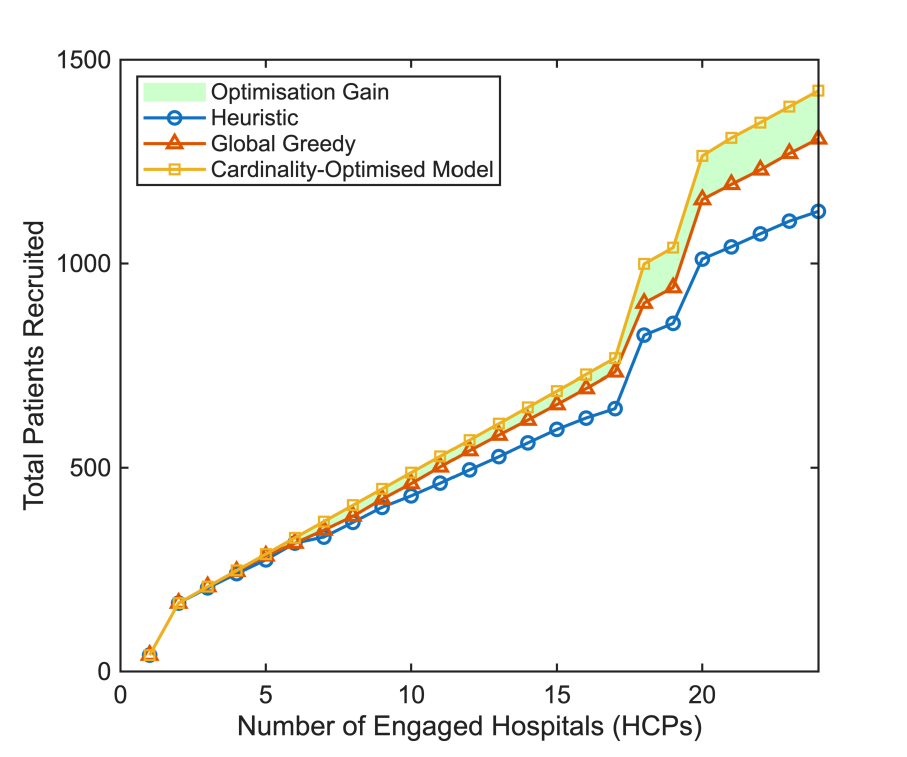}\caption{Optimisation gain vs heuristic (HCP engagement). Blue circles represent
the proportional heuristic recruitment strategy (Algorithm S1), red
triangles denote the global greedy recruitment baseline (Algorithm
S2), and yellow squares correspond to the cardinality-optimised model.
The shaded green region indicates the recruitment gain achieved by
the optimisation model relative to the global greedy recruitment baseline.
\protect\label{fig:modelA_Heuristic}}
\end{figure}

The results, shown in Figure \ref{fig:modelA_Heuristic}, indicate
that while all approaches performed similarly in single or two-centre
configurations, differences emerged as network size increased. The
proportional heuristic recruitment strategy exhibited limited ability
to exploit cross-site complementarities, leading to reduced recruitment
totals in larger networks. The greedy recruitment baseline demonstrated
improved coordination relative to the proportional heuristic; however,
it remained locally driven and did not achieve the same recruitment
levels as the cardinality-optimised model. As network complexity increased,
the performance gap between the optimisation model and both baseline
strategies widened, reflecting the growing importance of global coordination
across multiple diseases and hospitals.

\begin{figure}[H]
\centering
\includegraphics[scale=0.8]{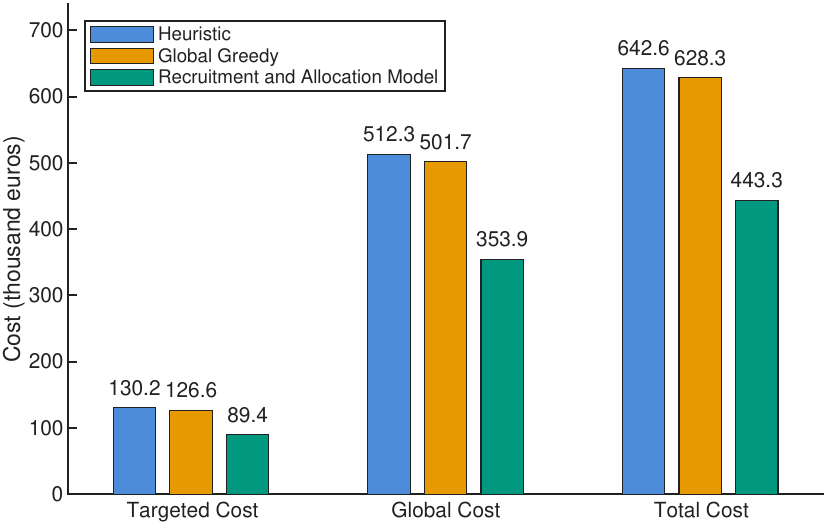}\caption{Metabolomic analysis cost comparison for 843 patients. Blue bars represent
the heuristic disease-wise allocation strategy (Algorithm S3), orange
bars denote the global greedy marginal allocation baseline (Algorithm
S4), and green bars correspond to the cardinality-optimised recruitment
and allocation model. Costs are reported in thousand euros and are
decomposed into targeted platform costs, global platform costs, and
total analysis cost. \protect\label{fig:metabolomic_cost_comparison}}
\end{figure}

Under unconstrained laboratory capacity conditions, the recruitment
and allocation model achieved approximately a 30\% reduction in total
metabolomic analysis cost compared to the heuristic disease-wise allocation
strategy (Algorithm S3). The global greedy marginal allocation baseline
(Algorithm S4) reduced costs relative to the heuristic approach but
remained more expensive than the cardinality-optimised solution. Cost
savings were observed for both targeted and global platforms. Figure
\ref{fig:metabolomic_cost_comparison} presents the comparative cost
breakdown across all three allocation strategies.

\begin{figure}[H]
\centering
\includegraphics[scale=0.5]{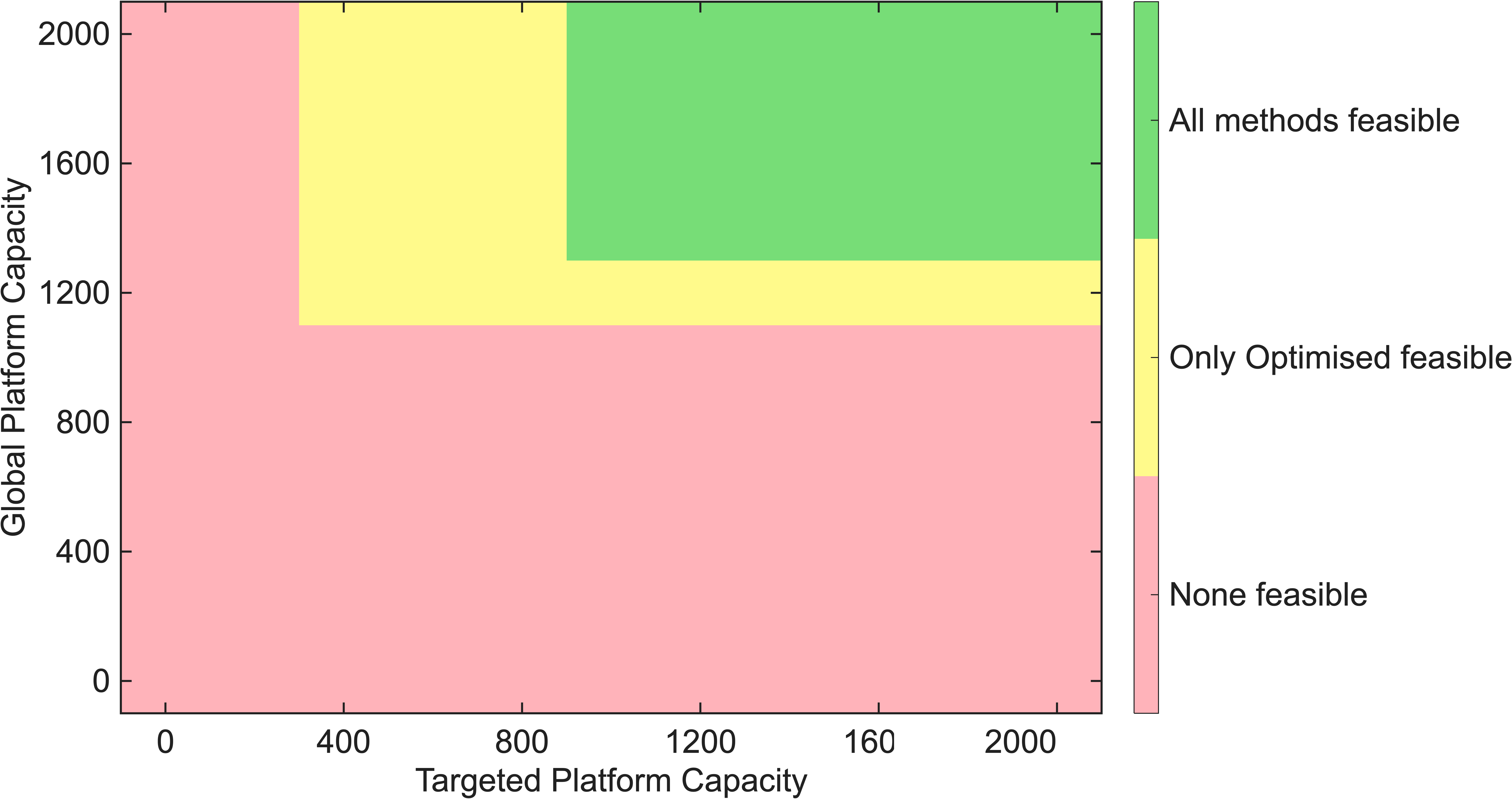}

\caption{Feasibility comparison of sample allocation strategies. The x-axis
and y-axis represent the maximum number of samples that can be analysed
by the targeted and global metabolomic platforms, respectively. Each
cell corresponds to a tested combination of platform capacity limits.
Green regions indicate configurations where the optimisation model,
the heuristic allocation strategy (Algorithm S3), and the greedy allocation
baseline (Algorithm S4) all return feasible solutions. Yellow regions
denote cases where only the cardinality-optimised model remains feasible,
while red regions correspond to scenarios in which none of the methods
can satisfy the coverage requirements. \protect\label{fig:Feasibility-comparison}}
\end{figure}

Figure \ref{fig:Feasibility-comparison} illustrates the feasibility
landscape across varying targeted and global platform capacity constraints.
Each grid cell represents a specific combination of platform throughput
limits, classified according to whether the optimisation model, the
heuristic allocation strategy (Algorithm S3), and the greedy allocation
baseline (Algorithm S4) are feasible. Green regions indicate configurations
where all approaches identify valid allocations, yellow regions denote
cases where only the cardinality-optimised model remains feasible,
and red regions correspond to scenarios where no method can satisfy
the coverage requirements. Notably, the heuristic and greedy baselines
became infeasible at identical capacity thresholds, whereas the optimisation
model remained feasible over a strictly larger region of the parameter
space. Across all tested capacity levels, there were no instances
in which either baseline method was feasible while the optimisation
model failed, highlighting the robustness of the proposed formulation
under progressively tighter operational constraints.

\subsection*{\protect\label{subsec:Trade-off-between-clinical}Trade-off between
clinical interest and logistic overhead}

\begin{figure}[H]
\centering
\includegraphics[scale=0.7]{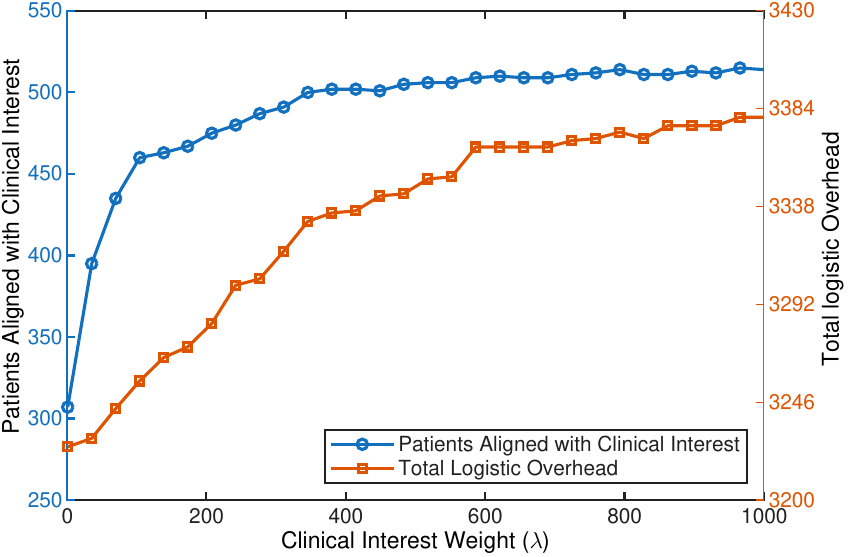}\caption{Clinical interest vs logistic overhead. Pareto frontier illustrating
the trade-off between two competing objectives: the number of recruited
patients aligned with hospital-specific clinical interests (left axis,
circles) and logistic overhead which is a summation of logistic deviation
variables (right axis, squares). Each point represents the solution
at a different value of the clinical interest coefficient ($\lambda$),
ranging from 1 to 1000. \protect\label{fig:clinical_interest_vs_logistics}}
\end{figure}

Figure \ref{fig:clinical_interest_vs_logistics} illustrates the trade-off
between clinical alignment and logistics cost. As the clinical interest
weight $\lambda$ increases, the recruitment pattern increasingly
reflects each hospital\textquoteright s interest in specific IMDs.
However, this improved alignment comes at the cost of reduced local
processing and increased inter-hospital sample shipments. The resulting
curve traces out a meaningful Pareto frontier, confirming the model\textquoteright s
ability to capture the inherent tension between clinical relevance
and logistical efficiency.

The curve shows a clear pattern of diminishing returns. Initial increases
in $\lambda$ produce marked gains in clinical-interest alignment,
whereas higher values produce smaller additional gains while logistics
overhead continues to rise. Using the Pareto-knee criterion described
in the Methods, $\lambda$ = 70 was selected for the primary analyses.
This value represents the point at which clinical-interest alignment
is incorporated without imposing disproportionate logistics overhead.

\section*{Discussion}

Despite common concerns about the computational cost of optimisation-based
methods \citep{davila_laboratory_2018}, our results suggest that,
with appropriate vectorisation, sparse reformulation, and solver selection,
such models can scale efficiently in large multi-centre study settings.
In the Recon4IMD case study, the model remained computationally tractable
as recruitment, allocation, capacity, cost, logistics, and clinical-interest
components were progressively incorporated. Supporting metrics, including
model dimensionality and solving times across model versions, are
provided in Supplementary Table S1. More generally, computational
burden is expected to increase with the number of diseases, healthcare
providers, biomarkers, and analytical platforms, since these determine
the size of the decision-variable vector and associated constraint
matrices. These considerations are important when adapting the framework
to larger rare disease networks.

Beyond efficiency gains, mathematical optimisation frameworks improve
fairness and transparency in operational decisions. In multi-site
studies like Recon4IMD, decisions regarding patient recruitment and
sample allocation are effectively a zero-sum game, where resources
such as patient access or analytical capacity at one site are allocated
at the expense of another. This creates potential for competing institutional
incentives and places considerable pressure and responsibility on
the project\textquoteright s coordination unit to make equitable decisions
under competing demands. Moreover, budget holders require clear justification
for how resources are distributed, particularly when public or pooled
funding is involved. An optimisation-based approach addresses these
concerns by offering a formally defined, reproducible, and data-driven
framework for decision-making. It enables the simultaneous consideration
of multiple factors (such as recruitment feasibility, analytical throughput,
cost, logistics, and clinical interest) that would be difficult for
any individual or committee to balance manually. In doing so, it not
only improves operational robustness but also promotes stakeholder
trust by providing transparent and justifiable allocation outcomes.

In multi-site clinical studies, optimisation-driven allocation must
also consider potential sources of systematic bias. Recruitment patterns
may unintentionally overrepresent certain geographic regions, demographic
groups, or clinical subpopulations, potentially conflicting with diversity,
equity, and inclusion objectives or regulatory expectations. Likewise,
heterogeneity in site-level performance, such as protocol deviation
rates, screen failure patterns, or competing trial involvement, may
introduce differences in data quality, and correlations between disease
distribution and laboratory assignment could confound biological signals
with platform effects. The present study did not include individual-level
demographic or site-quality data, and therefore, such fairness-aware
or bias-mitigation constraints were not incorporated into the optimisation
model. However, the linear constraint-based structure of the framework
readily permits extension through additional representation constraints,
site-level weighting factors, or platform-balancing requirements where
such data are available. For example, future implementations should
consider lower and upper bounds on recruitment by age group, sex,
geographic region, or under-represented subgroup, platform-balancing
constraints to avoid confounding biological subgroups with laboratory
assignment, and site-quality weights based on protocol-deviation or
screen-failure rates. Incorporating these elements would enable bias-aware
optimisation while preserving transparency and formal rigour in allocation
decisions.

While the optimisation-based framework offers considerable benefits
in terms of efficiency, fairness, and transparency, it is important
to acknowledge that model feasibility is not guaranteed under all
circumstances. Specifically, if the constraints imposed on the system
are mutually incompatible, the optimisation problem will have no solution.
For example, infeasibility can arise if the disease-specific cohort
size requirements, healthcare provider recruitment capacities, and
site-specific patient availability do not admit a distribution of
patients that simultaneously satisfies all constraints. This could
occur if certain diseases require more patients than are available
across the network, or if site-specific capacities are too restrictive
to permit allocation that meets cohort targets. In practice, recognising
and diagnosing such infeasibility is a key advantage of mathematical
optimisation. Empirically, our feasibility landscape analysis demonstrated
that the optimisation model remained feasible over a strictly larger
region of the operational parameter space than either baseline strategy.
Importantly, within the tested platform-capacity configurations and
under identical cohort definitions, no scenario was observed in which
either baseline method achieved feasibility while the optimisation
model failed. Accordingly, model outputs should always be interpreted
in light of the input data and constraints, and users are encouraged
to perform sensitivity analyses to identify constraint configurations
that preserve feasibility. In addition to structural infeasibility
arising from model constraints, practical implementation also faces
several real-world barriers related to patient availability and participation.
The integrality checks confirmed that all reported recruitment and
allocation quantities were integer-valued within solver tolerance
and therefore interpretable as patient and sample counts.

However, beyond strict feasibility considerations, the stability of
optimal allocations under changing assumptions also warrants attention.
Because the framework incorporates cardinality objectives, small perturbations
in input parameters, such as updated patient availability, capacity
limits, or biomarker mappings, may lead to discrete structural changes
in the optimal solution. This sensitivity is characteristic of non-convex
optimisation and may result in substantially different recruitment
or platform assignment patterns when assumptions are revised. In operational
settings, re-optimisation should therefore be performed with caution.
Rather than recalculating allocations from scratch, previously executed
or contractually committed decisions may be fixed as additional constraints,
preserving continuity of ongoing recruitment or laboratory processing.
The framework readily accommodates such constraints by restricting
selected decision variables during iterative re-optimisation. In this
way, adaptive planning can be achieved while maintaining operational
stability. Although extensions based on robust or stochastic optimisation
could explicitly model parameter uncertainty, such formulations would
substantially increase computational complexity and were beyond the
scope of the present study.

In addition to these structural considerations, real-world implementation
faces further barriers, particularly in rare disease contexts, where
patient availability, consent logistics, and participation rates may
fall short of expectations. To address this, we adjusted our model
inputs to reflect an estimated number of realistically recruitable
patients. This enhances the practical relevance of the optimisation
outcomes by aligning them with expected recruitment attrition rates
observed in real-world rare disease studies, thereby improving the
reliability of projected cohort feasibility.

Mathematical optimisation provides a structured basis for planning,
but its practical implementation in clinical research remains inherently
collaborative. In real-world settings, certain decisions, such as
adjusting to unforeseen staffing shortages, regulatory delays, or
evolving clinical priorities, may fall outside the scope of even the
most comprehensive models. Rather than replacing human oversight,
the optimisation framework is best understood as a decision support
tool that streamlines and formalises resource allocFation, while still
requiring refinement through expert judgement. This human-in-the-loop
approach enables research coordinators to adapt optimised plans to
live operational conditions, ensuring both robustness and contextual
appropriateness. By combining algorithmic rigour with clinical insight,
this strategy fosters more resilient and resource-efficient study
coordination. In the present analysis, robustness was assessed through
sensitivity analyses of HCP engagement, laboratory capacity limits,
and the clinical-interest weighting parameter. Future efforts should
continue to integrate dynamic operational data to further align optimisation
outputs with the evolving realities of clinical research.

This study demonstrates the value of a unified optimisation framework
for patient recruitment and sample allocation in complex, multi-site
rare disease studies. Applied to the Recon4IMD project, the approach
integrated cardinality optimisation, cost and logistics modelling,
and clinical alignment, enabling robust and efficient operational
planning across a European research network. Compared to proportional
heuristic and greedy baseline strategies, the optimisation model consistently
achieved higher recruitment yields, remained feasible under tighter
capacity constraints, and reduced overall metabolomic analysis costs.
The flexibility to incorporate multiple objectives and operational
constraints supports transparent and data-driven decision-making in
clinical research, offering a scalable solution for future large-scale,
resource-limited studies. As database resources, omics technologies,
and cross-border research infrastructures continue to evolve, optimisation-based
planning frameworks will be essential for coordinating increasingly
complex, multi-objective biomedical collaborations under real-world
resource constraints.

\section*{Methods}

This section outlines the mathematical framework used to model clinical
recruitment and metabolomic sample allocation. Central to our approach
is the use of cardinality optimisation, which enables the selection
of a minimal set of recruiting centres or resources while satisfying
estimated patient counts, recruitment capacity, and cohort targets.
We begin by formulating a general recruitment plan as a cardinality
optimisation problem, based on estimated patient count data, site-specific
recruitment capacity constraints, and disease-specific cohort recruitment
targets. Subsequent subsections expand the optimisation model to incorporate
sample allocation to analytical laboratories, cost-efficient platform
selection for omics analysis, logistical considerations, and alignment
with the clinical interests of healthcare providers. Finally, the
general framework is applied to the Recon4IMD project, with a detailed
explanation of how feasible IMDs were selected for investigation and
how the optimisation model was adapted to this case. The full mathematical
formulation, including the vectorised implementation, is provided
in the Supplementary sections 'Detailed Formulation of Sample Allocation
Optimisation in Recon4IMD' and 'Transforming optimisation formulas
for computational implementation'.

Notation convention: Input data matrices are denoted using bold capital
letters (e.g. $\mathbf{D}$ for the estimated patient count matrix).
Variable matrices are represented by capital italic letters (e.g.
$X,Y$), while scalar and vector variables are shown as lowercase
italic and Greek letters (e.g. $y,\:\beta$). While the same notation
is used throughout the general mathematical formulation and the Recon4IMD
case study, specific variables may correspond to particular datasets
or constraints in the applied context. A full summary of notation
is provided in Supplementary Table S3.

\subsection*{Clinical recruitment optimisation}

To enable effective patient recruitment across hospitals, the optimisation
framework begins with detailed hospital-level estimated patient count
data for each disease. These data represent the number of potentially
accessible patients at each site and serve as the empirical basis
for the recruitment model. They are used as input in the mathematical
formulation. The model links the estimated patient counts to recruitment
allocation decisions, such as how many patients to recruit per disease
at each hospital, while accounting for site-specific capacity and
feasibility constraints.

Let $\mathbf{D}\in\mathbb{N}^{m\times n}$ be \textit{\emph{an estimated
patient count matrix}} where $\mathbf{D}_{i,j}$ is the estimated
number of patients of disease $i$ at healthcare provider (HCP) $j$.
Let $X\in\mathbb{R}^{m\times n}$ be a \textit{\emph{recruitment fraction
matrix}} where $X_{i,j}\in[0,1]$ is the fraction of patients of disease
$i$ recruited at HCP $j$. Let $Y\in\mathbb{R}^{m\times n}$ be a
\textit{\emph{recruitment number matrix}} where $Y_{i,j}\in\mathbb{R}$
represents the planned number of patients with disease $i$ recruited
at HCP $j$. Then we have
\begin{equation}
\mathbf{D}\circ X=Y,\label{eq:DX=00003DY}
\end{equation}
where $\circ$ denotes the Hadamard product, that is, $\mathbf{D}_{i,j}\times X_{i,j}=Y_{i,j}$.
The recruitment numbers are modelled as continuous decision variables
representing planned allocation levels. Although patient counts are
inherently integer-valued, the optimisation problem is formulated
as a continuous relaxation to improve computational tractability within
the integrated recruitment and allocation framework. Because patient
and sample counts are discrete, final recruitment and allocation outputs
were checked for integrality by comparing each value with its nearest
integer within solver tolerance before reporting the solution. No
non-integer recruitment or allocation values were observed in the
reported analyses.

Minimising the number of active disease--HCP recruitment assignments,
thereby encouraging each disease to be recruited at fewer sites and
reducing fragmentation of recruitment efforts, while simultaneously
maximising the number of patients recruited per disease, can be formulated
as the following cardinality optimisation problem:
\begin{equation}
\begin{array}{cc}
\underset{X,Y}{\textrm{min}} & \left\Vert X\right\Vert _{0}-\mathds{\mathbf{1}}^{\mathrm{T}}y,\\
\textrm{s.t.} & \mathbf{D}\circ X-Y=0,\\
 & Y\mathds{\mathbf{1}}-y=0,\\
 & Y^{\mathrm{T}}\mathds{\mathbf{1}}-z=0,\\
 & 0\leq X\leq1,\\
 & \underline{\mathbf{y}}\le y\le\mathbf{\overline{y}},\\
 & \mathbf{\underline{z}}\le z\le\overline{\mathbf{z}}.
\end{array}\label{eq:minCard2_1}
\end{equation}
In this formulation, the term $\left\Vert X\right\Vert _{0}$ counts
the number of non-zero disease--HCP recruitment assignments, that
is, the number of disease--site pairs $(i,j)$ for which recruitment
is active. Minimising $\left\Vert X\right\Vert _{0}$, therefore,
encourages recruitment for each disease to be concentrated across
fewer sites, reducing fragmentation of recruitment activities. Operationally,
such concentration can reduce administrative and monitoring burden,
simplify coordination and contracting across centres, and support
more consistent protocol implementation and data collection. From
a statistical perspective, limiting fragmentation may mitigate inter-site
heterogeneity in clinical practice and measurement procedures, thereby
improving comparability of observations across the network.

The vector $y\in\mathbb{R}^{m}$ represents the planned total number
of patients recruited per disease, and $z$$\in\mathbb{R}^{n}$ represents
the planned total number of patients recruited by each healthcare
provider. Here, $\mathds{\mathbf{1}}$ denotes a vector of ones with
appropriate dimension. The lower and upper bounds $\underline{\mathbf{y}}$
and $\mathrm{\mathbf{\overline{y}}}$ constrain the number of recruited
patients for each disease, based on each disease cohort. Similarly,
$\underline{\mathbf{z}}$ and $\mathrm{\mathbf{\overline{z}}}$ specify
the recruitment capacity limits for each healthcare provider, reflecting
constraints such as available staff, or infrastructure.

\subsection*{Biomarker demand induced by patient recruitment}

As patient samples are destined for omics analysis, the patient recruitment
and sample allocation strategies must be designed in an integrated
manner. Because samples correspond to different inherited metabolic
disorders (IMDs), the set of biomarkers to be measured varies by disease.
The relationships between diseases and their characteristic biomarkers
can be obtained from established domain-specific resources, such as
IEMbase \citep{ferreira_proposed_2019}, as well as from curated clinical
literature. These disease--biomarker associations provide the biological
foundation for determining measurement requirements within the optimisation
framework.

Let $\mathbf{B}\in\{0,1\}{}^{p\times m}$ be a \textit{\emph{biomarker-association}}\textit{
}matrix, where $\mathbf{B}_{i,j}=1$ if biomarker $i$ is associated
with disease $j$, and $\mathbf{B}_{i,j}=0$ otherwise. Let $y\in\mathbb{R}^{m}$
denote the planned total number of patients recruited per disease.
The resulting biomarker measurement matrix is defined as:

\begin{equation}
S=\mathbf{B}\textrm{diag}\left(y\right),\label{eq:By=00003DS}
\end{equation}

where $S\in\mathbb{R}^{p\times m}$ and $S_{i,j}$ represents the
number of times biomarker $i$ must be measured for disease $j$.
This formulation ensures that biomarker demand scales proportionally
with disease-specific recruitment levels and establishes a direct
link between patient recruitment and downstream analytical requirements.

\subsection*{Laboratory allocation and capacity constraints}

Designing an efficient omics analysis strategy involves determining
a consolidated allocation of analytical platforms required to satisfy
the biomarker measurement matrix $S$, while respecting laboratory
capacity and operational constraints. In addition to coverage considerations,
the allocation must reflect the analytical capabilities and throughput
limitations of each participating laboratory.

To mathematically model platform capabilities, data are collected
from participating laboratories specifying which biomarkers, such
as genes, proteins, or metabolites, are detectable by each analytical
test panel. These platform-specific biomarker lists are cross-mapped
and harmonised across laboratories to ensure consistent identifier
usage and enable fair comparison of coverage. Once harmonised, this
capability information serves as structured input to the optimisation
framework.

Let $\mathbf{T}\in\{0,1\}{}^{p\times q}$ be a \textit{\emph{platform
capability matrix}} where $\mathbf{T}_{i,j}=1$ if biomarker $i$
is detectable by analytical platform $j$ and $\mathbf{T}_{i,j}=0$
otherwise. Let $U\in\mathbb{R}^{q\times m}$ represent\textit{\emph{
platform utilisation}} where $U_{i,j}$ is the number of samples from
disease $j$ assigned to platform $i$.

To ensure that all required biomarkers are measured, the allocation
must satisfy:

\begin{equation}
\mathbf{T}U\geq S.\label{eq:TUgeS}
\end{equation}

The inequality allows biomarkers to be measured more than once where
necessary, providing flexibility in platform assignment.

To reduce fragmentation of disease-platform assignments while ensuring
full biomarker coverage, we formulate the following cardinality optimisation
problem:

\begin{eqnarray}
\underset{U}{\textrm{min}} & \left\Vert U\right\Vert _{0},\nonumber \\
\textrm{s.t.} & \mathbf{T}U\geq S,\label{eq:minTU=00003DS}
\end{eqnarray}

Because $S$ is defined in Eq. \ref{eq:By=00003DS} as a function
of the planned recruitment vector $y$, any non-zero recruitment implies
$S\neq0$, and therefore the trivial solution $U=0$ is infeasible.
The term $\left\Vert U\right\Vert _{0}$ counts the number of non-zero
disease--platform assignments, that is, the number of disease-specific
sample flows routed to analytical platforms. Minimising $\left\Vert U\right\Vert _{0}$
therefore promotes consolidation of analytical workflows, reducing
fragmentation of disease-specific measurements across multiple platforms.
Operationally, limiting cross-platform splitting can simplify logistics,
reduce sample handling complexity, and streamline coordination between
laboratories. Thus, $\left\Vert U\right\Vert _{0}$ is interpreted
as a sparsity penalty on individual disease--platform assignments.

This objective promotes assignment-level sparsity. Introducing explicit
platform-level binary selection variables would lead to a substantially
more complex mixed-integer formulation and was therefore avoided to
preserve computational tractability. The present formulation balances
analytical consolidation with computational tractability while maintaining
flexibility for platform reuse across multiple diseases.

To reflect real-world operational limits, laboratory throughput constraints
are also introduced. Each analytical platform is assumed to have a
finite processing capacity, restricting the total number of patient
samples that can be analysed by that platform. To model this, an auxiliary
variable is defined. Let $u\in\mathbb{R^{\mathit{q}}}$ be the total
number of samples allocated to each analytical platform, where $u_{i}$
represents the total number of samples assigned to analytical platform
$i$. This auxiliary variable is linked to the platform utilisation
matrix through the relation:

\begin{equation}
u=U\mathds{\mathbf{1}}_{m},
\end{equation}
where $\mathds{\mathbf{1}}_{m}$ denotes a column vector of ones of
length $m,$ corresponding to the number of diseases.

Capacity limits are imposed as:

\begin{equation}
\mathrm{\begin{array}{cc}
0\leq\mathit{u_{i}}\leq\mathbf{\overline{u}_{\mathit{i}}}, & \forall\:i=1,\ldots,\mathit{q},\end{array}}
\end{equation}

where $\mathbf{\overline{u}_{\mathit{i}}}$ is the maximum number
of samples that analytical platform $i$ is capable of processing.

This formulation ensures that biomarker demand induced by recruitment
is satisfied, while reducing disease-platform allocation fragmentation
and respecting operational capacity constraints.

\subsection*{Cost and logistics modelling}

In many multi-centre omics studies, multiple analytical laboratories
participate, each offering different platforms, operating under distinct
technical and logistical conditions. The cost of omics analysis is
therefore, both laboratory-specific and platform-dependent. Additional
factors such as the number of samples, batch sizes, and sample types
further influence the overall cost of analysis. As with most clinical
research projects, budget constraints play a critical role in shaping
sample allocation strategies. Consequently, mathematical modelling
of sample allocation should incorporate the associated analytical
costs to ensure that the selected strategy remains within budget.

Let $\mathbf{r}\in\mathbb{R}^{1\times q}$ denote the platform cost
vector, where $\mathbf{r}_{i}$ represents the cost of performing
one omics analysis using platform $i$. Let $\beta\in\mathbb{R}^{1\times m}$
represent the total omics analysis cost per disease, where $\beta_{j}$
corresponds to the overall cost associated with analysing samples
from disease $j$. The relationship between platform utilisation and
disease-specific costs is defined as:

\begin{equation}
\mathbf{r}U=\beta.\label{eq:targetedPlatformCost}
\end{equation}

This formulation assumes a linear cost model, where total costs scale
directly with the number of assigned analyses, and does not account
for potential non-linearities such as batch effects, minimum order
sizes, or setup overhead.

\subsection*{Incorporation of clinical interest in the recruitment strategy}

In addition to the logistical and operational factors previously considered
in the recruitment strategy, the clinical interests of each healthcare
provider also play an important role. Hospitals with focused expertise
in certain diseases are more likely to have experienced staff familiar
with the associated diagnostic, treatment, and management requirements.
These sites may also be involved in ongoing research or clinical studies
related to specific diseases, thereby increasing the likelihood of
successful patient recruitment and long-term engagement.

To incorporate site-specific clinical expertise into the recruitment
optimisation framework, a linear objective coefficient $c\in\mathbb{R}^{d}$
is introduced, where $d$ denotes the total number of decision variables
in the vectorised optimisation problem. Specifically, to reflect the
clinical interest of each healthcare provider, the coefficient $c$
is assigned to each disease-hospital pair $(i,j)$ and reflect as
\begin{equation}
c(i,j)=-\lambda\:\times\:\mathbf{C}(i,j),\label{eq:clinical_interest}
\end{equation}

where $\mathbf{C}\in\mathbb{R}^{m\times n}$ is the clinical interest
matrix, with $\mathrm{\mathbf{C}(\mathit{i,j})}\:=\:1$ if hospital
$j$ is interested in recruiting patients with disease $i$, and 0
otherwise. $\lambda$ referred to as the \textit{\emph{clinical interest
weight}}, is a user-defined parameter that reflects the relative importance
of clinical interest compared to other objectives in the model. Since
the optimisation problem is formulated as a minimisation, the negative
sign ensures that clinical interest contributes positively by reducing
the overall objective value, thus encouraging the allocation of patients
to hospitals aligned with specific disease interests. The parameter
$\lambda$, controls the strength of this incentive relative to other
objectives in the model.

The selection of $\lambda$ was calibrated through a sensitivity analysis
over the tested range of clinical-interest weights. For each value
of $\lambda$, we recorded two outcomes: the number of recruited patients
aligned with declared clinical interest, denoted $\mathrm{A}(\lambda)$,
and the total logistics deviation, denoted $\mathrm{L}(\lambda)$.
Both quantities were min-max normalised to $[0,1]$. Since $\mathrm{A}(\lambda)$
is maximised and $\mathrm{L}(\lambda)$ is minimised, the selected
value of \textgreek{\textlambda} was defined using a Pareto-knee criterion.
Specifically, the selected value was the point with the maximum signed
perpendicular distance from the chord connecting the two extreme normalised
Pareto solutions. This identifies the point at which further improvement
in clinical-interest alignment requires a disproportionate increase
in logistics deviation. This approach follows established knee-point
selection principles for bicriteria Pareto trade-offs \citep{deb_understanding_2011}.

It is important to note that this clinical interest term acts as a
soft objective and not a constraint. The optimisation always ensures
feasibility based on estimated patient counts and site capacity. When
multiple hospitals have sufficient patient populations for a given
disease, the model will prioritise assignments to those that have
declared clinical interest. If only one among the eligible hospitals
expresses interest, that site will be preferred. In cases where no
hospital declares interest for a disease, the allocation will still
proceed based on feasibility. Thus, the clinical interest term provides
a directional preference that guides allocation decisions when alternatives
exist.

\subsection*{Modelling sample allocation logistics}

In many real-world clinical studies, certain healthcare providers
are either directly equipped with in-house omics laboratories or maintain
collaborative agreements with nearby analytical facilities. These
hybrid sites can both recruit patients and perform omics analyses
locally. In such scenarios, it may be inefficient or unnecessary to
transport biological samples over long distances when local analysis
is feasible. To reflect this consideration, the objective function
can be augmented with a penalty for allocating samples to non-local
or logistically distant laboratories.

Let $Y\in\mathbb{R}^{m\times t}$ denote a subset of the recruitment
matrix $Y$ , where $t$ is the number of hybrid or co-located sites
equipped with omics platforms. In an ideal scenario, there would be
full alignment between the number of patients recruited at these sites
and the corresponding platform utilisation, such that samples collected
at a given site are analysed locally. This can be expressed as:

\begin{equation}
\begin{array}{cc}
Y_{i,j}=U_{i,j}^{\mathrm{T}}, & \forall\:i=1,\ldots,m,j=1,\ldots,t.\end{array}\label{eq:Ysub_equals_Usub}
\end{equation}

Deviations from ideal local processing are captured through a \textit{\emph{logistics
deviation matrix}} $F_{i,j}\in\mathbb{R}^{m\times t}$, defined as:

\begin{equation}
\begin{array}{cc}
F_{i,j}\coloneqq\mid Y_{i,j}-U_{i,j}^{\mathrm{T}}\mid & \forall\:i=1,\ldots,m,j=1,\ldots,t.\end{array}\label{eq:logistic_deviation}
\end{equation}

Here, $F_{i,j}=0$ indicates that samples collected at site $j$ for
disease $i$ are analysed locally, with no external shipment required.
Since such perfect alignment is rarely achievable in practice, this
constraint is relaxed:

\begin{equation}
F_{i,j}\geq\mid Y_{i,j}-U_{i,j}^{\mathrm{T}}\mid.
\end{equation}
To discourage unnecessary sample shipment and promote local analysis
where possible, the model includes a penalty term in the objective
function that minimises the total logistical deviation. This is expressed
as the one-norm of the deviation matrix:
\begin{equation}
\parallel F\parallel_{1}=\sum_{i=1}^{m}\sum_{j=1}^{t}F_{i,j}.
\end{equation}

Minimising this term supports cost reduction, improves logistical
efficiency, and enhances overall robustness of the sample allocation
strategy.

\subsection*{Scaling of Objective Components}

The integrated objective function combines heterogeneous components,
including operational limitations, recruitment totals, logistics deviations,
and monetary cost terms. Because these components are expressed in
different natural units, their numerical magnitudes may differ substantially.
In weighted-sum formulations, components operating at higher orders
of magnitude may exert disproportionate influence unless explicitly
scaled.

Structural terms such as $\parallel X\parallel_{0},$ $\parallel U\parallel_{0},$
recruitment totals $\parallel y\parallel_{1},$ and logistics deviations
$\parallel F\parallel_{1},$ $\parallel W\parallel_{1},$ are bounded
by the number of diseases, sites, platforms, and cohort limits, typically
operating at magnitudes on the order of $10^{2}-10^{3}$. In contrast,
laboratory cost terms scale with per-sample platform costs and total
cohort size, and may operate at magnitudes several orders higher.
For instance, assuming an estimated per-sample omics analysis cost
of 100 euros, a cohort of approximately 1,000 patients would yield
a total laboratory cost $\parallel\beta\parallel_{1}$ to be on the
order of $10^{5}$, which is two orders of magnitude larger than operational
components driven by patient or sample counts.

To ensure comparable numerical magnitudes across objective components
while preserving interpretability in natural units, a scaling coefficient
$\rho\in\mathbb{R>\mathrm{0}}$ is introduced for cost-related $l_{1}$
terms. The resulting objective function can be written as:

\begin{equation}
\mathrm{min}\quad c^{\mathrm{T}}\varphi(v)+\parallel X\parallel_{0}-\parallel y\parallel_{1}+\parallel U\parallel_{0}+\parallel F\parallel_{1}+\rho\parallel\beta\parallel_{1},
\end{equation}

where $\rho$ controls the numerical scale of monetary cost contributions
relative to operational and logistical components. In this formulation,
the vector $v$ contains all decision variables of the model, and
$\varphi(v)\in\mathbb{R^{\mathrm{\mathit{d}}}}$ denotes the vectorisation
operator that maps these structured variables into a single decision
vector compatible with the objective coefficient vector $c\in\mathbb{R^{\mathrm{\mathit{d}}}}$.

The $\ell_{0}$-type terms in the objective, including $\parallel X\parallel_{0}$
and $\parallel U\parallel_{0}$, were handled through an established
cardinality optimisation implementation \citep{fleming_cardinality_2023a,heirendt_creation_2019c}.
In this approach, the zero-norm is approximated using non-convex continuous
functions, reformulated as a difference-of-convex optimisation problem,
and solved approximately through a sequence of convex optimisation
subproblems. Gurobi was used as the solver for these subproblems \citep{gurobioptimizationllc_gurobi_2024}.
The implemented model is therefore a cardinality approximation based
on a continuous reformulation, rather than an exact mixed-integer
or exhaustive enumeration formulation.

Solver diagnostics were recorded for the reported optimisation model
variants. Because the zero-norm terms were handled through a continuous
cardinality approximation rather than an exact mixed-integer formulation,
MIP gaps are not applicable. For each reported model variant, Gurobi
returned optimal termination for the implemented optimisation problem,
and the final solution satisfied the implemented constraints with
zero maximum feasibility residual under the diagnostic checks. These
diagnostics, together with the existing model dimensionality and solving-time
information, are reported in Supplementary Tables S1 and S2. Accordingly,
the reported solutions should be interpreted as feasible approximate
cardinality-optimised solutions, rather than certified global optima
of the original zero-norm formulation.

The cardinality optimisation framework supports such scaling via user-defined
objective coefficients applied to selected entries of the vectorised
decision vector $\varphi(v)$, allowing controlled adjustment of relative
influence without modifying the constraint structure \citep{fleming_cardinality_2023a}.
In the Recon4IMD implementation, we set $\rho=10^{-2}$, chosen to
align the numerical magnitude of cost terms with operational components.

\subsection*{Unified optimisation formulation\protect\label{subsec:Unified-optimisation}}

\begin{figure}[H]
\centering
\includegraphics[scale=1.1]{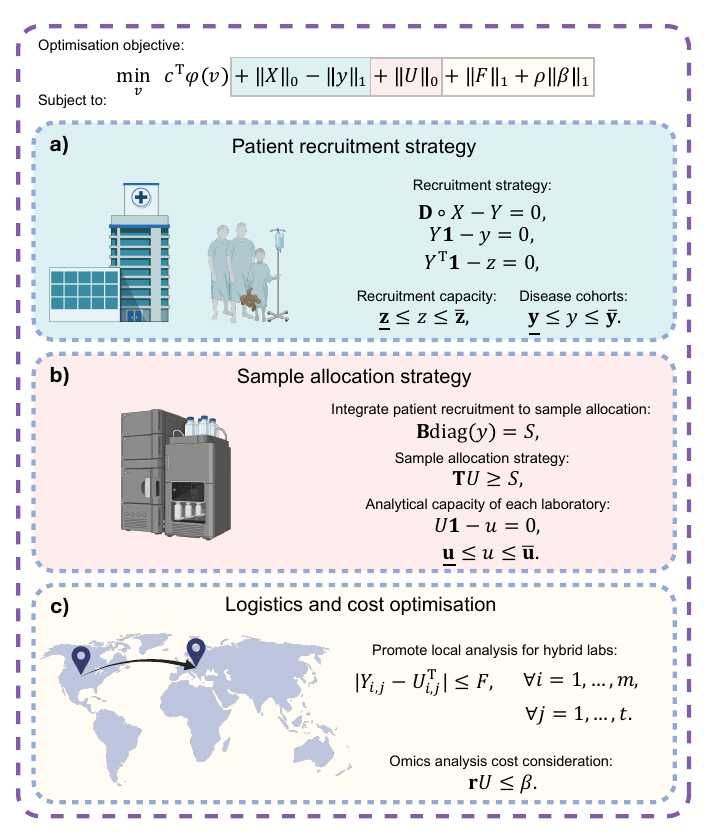}\caption{Unified optimisation framework for multi-centre clinical study design.
This framework integrates clinical recruitment, sample allocation,
and logistical constraints into a single mathematical model. a) Patient
recruitment strategy: selecting healthcare providers to recruit patients
based on disease-specific estimated patient counts at each site, target
disease cohorts, and site-specific capacities. b) Sample allocation
strategy: assigning biological samples to analytical platforms while
ensuring full biomarker coverage and laboratory throughput constraints.
c) Logistics and cost optimisation: minimising non-local sample shipments
and analytical costs. The objective function incorporates clinical
interest through linear objective coefficients and combines cardinality
and one-norm terms to balance recruitment, platform usage, logistics,
and cost efficiency.\protect\label{fig:unified-optimisation}}
\end{figure}

All previously discussed components, patient recruitment strategy,
omics sample distribution, cost minimisation, logistical deviation,
and clinical interest alignment, are integrated into a unified optimisation
framework. The resulting model balances multiple objectives, including
maximising the number of patients recruited across all diseases, minimising
the number of recruiting sites and platforms, encouraging local processing,
reducing costs, and aligning recruitment with clinical expertise.
The complete formulation is illustrated in Figure \ref{fig:unified-optimisation}.

\subsection*{Case study: Recon4IMD implementation}

Reconstruction and Computational Modelling for Inherited Metabolic
Diseases (Recon4IMD) is a pan-European research initiative aimed at
enhancing the diagnosis of IMDs through integrated metabolomic and
computational modelling approaches. IMDs comprise over 1,990 rare
genetic disorders with diverse and often ambiguous clinical presentations,
which complicate diagnosis and delay treatment. Given the limited
patient populations per site and the need for advanced analytical
workflows, Recon4IMD provides a representative case for optimising
recruitment and sample allocation across a distributed network.

To address these challenges, the Recon4IMD project mobilises a network
of healthcare providers across Europe to recruit diagnosed IMD patients,
with a total of 24 participating centres. Recruitment is constrained
by hospital-specific capacities and site-specific patient availability,
making efficient cohort assembly non-trivial. In parallel, urine and
plasma samples are collected and analysed at eight specialised metabolomics
laboratories with differing platform capabilities, throughput, and
costs. These analyses combine targeted and global metabolomics to
characterise disease-specific metabolic profiles and support the reconstruction
of IMD-specific genome-scale metabolic models.

The complexity and scale of this collaborative effort highlight the
operational interdependencies that motivate the unified optimisation
framework described in this study. By incorporating constraints on
recruitment feasibility, laboratory capacity, biomarker coverage,
cost, and clinical interest, the model ensures scalable, transparent,
and clinically aligned study design across the network.

\subsection*{Selection of IMDs in the Recon4IMD project}

The first step in the Recon4IMD recruitment strategy was to define
a feasible list of inherited metabolic diseases for inclusion. Given
the multidisciplinary scope of the project, selected IMDs needed to
meet criteria spanning computational modelling, analytical feasibility,
and clinical relevance. Each IMD had to have a known causal gene and
a corresponding metabolic pathway represented in the latest genome-scale
model of the human cell. The Virtual Metabolic Human (VMH) database
was used to identify relevant genes and metabolites, excluding IMDs
without clear genetic or metabolic biomarkers \citep{noronha_virtual_2019}.
Metadata, including causal genes, biomarkers, and age-specific effect
sizes, was obtained from the IEMbase database \citep{ferreira_proposed_2019}.

To ensure analytical feasibility, each participating metabolomics
laboratory provided a list of metabolites measurable by their platforms.
These were intersected with (i) IMD-associated biomarkers and (ii)
metabolites present in the genome-scale model. This mapping was performed
using MetaNetX, which integrates metabolic networks and biochemical
pathway databases under the unified MNXref namespace \citep{moretti_metanetx_2021}.

Finally, recruitment feasibility was addressed using hospital-provided
estimates of patient counts, ensuring that the selected IMDs had sufficient
patient numbers to support statistically robust clinical studies.

\subsection*{Healthcare providers in Recon4IMD project}

Healthcare providers (HCPs) serve as centres for recruiting patients
with IMDs. Each HCP has a limited number of patients available per
IMD and operates under a recruitment capacity determined by its facilities
and human resources. In this case, most hospitals have a recruitment
threshold of 40 patients, while three participating centres can recruit
up to 250 patients. These constraints define the allowable values
of the recruitment variable $z$., which represents the number of
patients recruited per HCP.

A survey was conducted to collect clinical interest from each HCP.
Each hospital submitted its preferred IMDs using IEMbase nosology
code as the namespace or a whole ICIMD nosology group \citep{ferreira_international_2021}.
By mapping ICIMD nosology groups to their associated IEMbase IMDs,
a clinical interest vector was constructed for each HCP at the IMD
level. This clinical interest vector is incorporated into the integrated
optimisation problem, as shown in Equation \ref{eq:clinical_interest}.

\subsection*{Effect size and cohort design by IMD}

To date, no systematic study has established how many patients are
needed to construct patient-specific genome-scale models across a
large number of IMDs. From an optimisation perspective, defining quantified
lower and upper bounds for cohort sizes is essential to balance statistical
power, diagnostic certainty, and resource constraints. To address
this gap, a scalable framework was developed to estimate appropriate
cohort sizes using metabolomic data as a primary source of diagnostic
information. This approach is intended as a rational starting point
for guiding recruitment, particularly in the absence of disease-specific
modelling precedents.

Diagnostic certainty is one of the key factors influencing the number
of patients required for a valid clinical study. Studies of diseases
that present clear metabolite biomarkers with a large effect size
generally require fewer patients \citep{burmeister_sample_2012,dobbin_sample_2022,serdar_sample_2021}.
Conversely, conditions with less distinctive biomarkers and smaller
effect sizes typically need larger cohorts to guard against misdiagnosis.

To formalise this relationship, we introduce the Diagnostic Confidence
Score (DCS), a quantitative metric designed to estimate diagnostic
reliability specifically from metabolomic data. The DCS integrates
two components: the number of characteristic metabolite biomarkers
and their respective effect sizes, as reported in IEMbase. This score
reflects how distinguishable a disease is based on its metabolomic
profile. Diseases with higher DCS values are assumed to be more confidently
identifiable from metabolomic data alone and are therefore expected
to require fewer patients for metabolic model construction. Conversely,
lower DCS values indicate limited or ambiguous metabolomic signals
and suggest a need for larger cohorts to mitigate diagnostic uncertainty.

The DCS is calculated using qualitative diagnostic annotations from
the IEMbase database. Effect sizes are encoded in IEMbase using arrows,
which are converted into numerical values to enable quantitative analysis:
one arrow is assigned a value of one unit, two arrows a value of two
units, and ambiguous ranges such as \textquotedblleft normal to one
arrow\textquotedblright{} are approximated as 0.5 units. The average
of these values across all age-specific groups defines the quantified
effect size for each IMD.

In parallel, we identify the number of feasible characteristic metabolite
biomarkers, defined as those both reported in IEMbase and measurable
by Recon4IMD partner laboratories. The Diagnostic Confidence Score
is then defined as:

{\footnotesize
\begin{equation}
\mathrm{Diagnostic\;Confidence\;Score\;=\;\left[number\;of\;feasible\;characteristic\;metabolite\;biomarkers\right]\;\times\;quantified\;effect\;size}
\end{equation}
}{\footnotesize\par}

This score serves as a proxy for the diagnostic utility of metabolomic
data in identifying a given IMD. To implement this framework in Recon4IMD,
we stratified IMDs into three diagnostic confidence categories based
on the DCS distribution: high (DCS $>$ 6), moderate (3 $\leq$ DCS
$\leq$ 6), and low (DCS $<$ 3). These thresholds were chosen based
on the empirical distribution of DCS values calculated for all IMDs
in IEMbase, as detailed in Supplementary Figure S1 and the accompanying
explanation. The distribution is highly left-skewed, with a substantial
proportion of IMDs exhibiting a DCS of zero, indicating no currently
measurable characteristic biomarkers.

The cohort size for each IMD is case-specific, making it difficult
to determine quantitatively the exact number of samples required for
a clinical study. A recent study confirmed that sample sizes in rare
disease studies vary widely, ranging from a few individuals to several
hundred \citep{hee_does_2017}. Several factors influence the determination
of an appropriate sample size, including the expected effect size
or difference between groups or time points, the degree of homogeneity
among study participants, the acceptable risk of error, and the anticipated
rate of participant attrition \citep{burmeister_sample_2012}. In
addition to these, the overall scope of the project and the availability
of resources such as funding, time, and personnel also play a significant
role in shaping sample size decisions, as they can limit or expand
the capacity to recruit and manage participants effectively. In this
study, we projected an average of 10--15 patients for each disease
cohort, which aligns with the Recon4IMD project plan to recruit approximately
1,900 diagnosed patients from a diverse group of IMDs. Initially,
the defined DCS thresholds correspond to natural breaks in the distribution
and were used to guide cohort size assignments: 5--10 patients for
high-confidence IMDs, 10--15 patients for moderate-confidence IMDs,
and 15--20 patients for low-confidence IMDs. Finally, the resulting
cohort size recommendations were reviewed by clinicians involved in
the project and adjusted on a case-by-case basis based on expert evaluation
of each disease. This review step also reflects that the Diagnostic
Confidence Score was used as a project-specific planning tool to guide
cohort-size bounds, rather than as a validated diagnostic score or
universal sample-size rule.

\subsection*{Adjusting patient estimates for practical recruitment limitations}

Recruitment for clinical studies is often limited by operational and
ethical challenges. Factors such as limited patient availability,
infrequent clinic attendance, and inconsistent collection of diagnostic
data can hinder the timely identification and enrolment of eligible
participants. Additional barriers include those specific to paediatric
studies (e.g. the need for dual parental consent in some countries),
patient cancellations, reluctance to participate, and, in some cases,
intentional avoidance of approach due to concerns about overburdening
patients or jeopardising the clinician--patient relationship. These
constraints reduce the realistically recruitable patient pool, even
when a larger number may be theoretically eligible. To improve alignment
with these limitations, we propose adjusting the model inputs to reflect
not the total recorded patient counts at each site, but an estimated
number of patients likely to participate. This modification increases
the practical applicability of the optimisation framework by incorporating
expected recruitment attrition into the planning assumptions, thereby
enhancing both the feasibility and robustness of cohort allocation
strategies.

\textit{\emph{Based on our experience, we anticipated that approximately
50\% of estimated patients would ultimately provide a sample for the
study. Therefore, we divided the estimated patient count matrix}}
$\mathbf{D}$ by two to more accurately approximate the number of
patients expected to be recruited. This adjustment can reduce the
estimated recruitable pool below the nominal disease-specific cohort
lower bound for some IMDs, making the integrated optimisation problem
infeasible if the original cohort thresholds are imposed unchanged.
We therefore applied a feasibility relaxation to the disease-specific
cohort lower bounds, defining the minimum acceptable cohort size as
a fixed fraction of the nominal target and rounding this value upward
to the nearest integer. This ensured that the relaxation did not introduce
fractional patient requirements. This approximation may vary across
hospitals, depending on their registration procedures, the specific
diseases involved, and the time span over which patient data are collected,
as older records typically correspond to lower participation rates.

\subsection*{Metabolomic analysis in Recon4IMD project}

Metabolomics is central to the Recon4IMD project, as it captures the
biochemical consequences of metabolic dysfunction in IMDs. Both quantitative
targeted and semi-quantitative global metabolomics are applied to
each patient sample to ensure comprehensive biochemical characterisation,
enabling integration with genome-scale metabolic models. Accordingly,
the mathematical modelling of sample allocation, as described in the
Methods section on sample allocation and logistics, applies to both
global and targeted platforms within Recon4IMD.

Seven participating laboratories are co-located with clinical centres,
forming hybrid sites capable of both patient recruitment and local
sample analysis. This distributed configuration introduces logistical
complexity into the sample allocation process, which must account
for platform capabilities, analytical throughput, cost, and geographic
proximity.

A detailed mathematical formulation of the integrated optimisation
strategy is provided in Supplementary Methods. This includes platform
capabilities, capacity constraints, cost modelling, and logistical
penalties.

\subsection*{Metabolite identifier standardisation and interdisciplinary integration}

The Recon4IMD project evaluated eight European metabolomic laboratories
with varying analytical platforms, costs, and sample throughput capacities,
highlighting the need for a unified metabolite identifier namespace
to objectively compare platform capabilities and integrate metabolomics
data into genome-scale models. Bridging experimental results with
computational models requires harmonising identifiers used in analytical
platforms with those in metabolic reconstructions. Multiple identifier
systems exist for different use cases: the Human Metabolome Database
(HMDB) offers human-centric annotations \citep{wishart_hmdb_2021};
KEGG Compound IDs support pathway mapping \citep{kanehisa_kegg_2023};
PubChem CIDs provide extensive structural data \citep{kim_pubchem_2023};
SwissLipids offers lipid-specific, ontology-based annotations \citep{aimo_swisslipids_2015a};
and ChEBI for ontology-driven annotation of chemical entities \citep{hastings_chebi_2016}.
On the modelling side, the Virtual Metabolic Human (VMH) database
serves as the reference for genome-scale reconstructions, linking
its identifiers to databases like ChEBI, KEGG, and HMDB for interoperability
\citep{noronha_virtual_2019}. Among these, ChEBI is notable for its
structured ontology and its explicit representation of stereochemical
and charged states, a critical feature given the limitations of many
metabolomics platforms in resolving stereoisomers or isomeric lipids,
especially in untargeted workflows \citep{oliveira_enantioselectivity_2021,addepalli_concise_2020}.
While stereoisomer identity can sometimes be inferred from biological
context, such as the predominance of L-amino acids in plasma, multiple
isomers often coexist, and ChEBI\textquoteright s hierarchical ontology
supports formal representation of this uncertainty \citep{novoa-del-toro_navigating_2024}.
Thus, ChEBI offers a robust, ontology-driven framework for metabolite
identification in projects like Recon4IMD, enabling precise, semantically
interoperable communication between experimental and computational
teams \citep{poupin_improving_2020}.

\subsection*{Data privacy and anonymisation strategy}

To comply with the General Data Protection Regulation (GDPR; Regulation
(EU) 2016/679) and ethical obligations around patient confidentiality,
we anonymised all sensitive data prior to analysis. Specifically,
we withheld the estimated number of patients recruited at each healthcare
provider and the allocation of samples to individual metabolomic laboratories.
This step was taken to prevent indirect re-identification of patient
subgroups or institutional profiles based on operational data.

To preserve analytical utility while ensuring confidentiality, we
evaluated several anonymisation techniques, including k-ananymity
\citep{sweeney_k-anonymity_2002}, data aggregation, and synthetic
data generation. We selected differential privacy as the most scientifically
rigorous and formally grounded approach \citep{dwork_calibrating_2006},
offering provable privacy guarantees by injecting calibrated statistical
noise. A privacy budget of $\varepsilon=1.5$ was used, balancing
privacy protection and model fidelity aligning with established practice
for healthcare and public policy applications \citep{dwork_algorithmic_2014,erlingsson_rappor_2014c}.

Further technical details, including the mathematical formulation
and implementation of the Laplace mechanism, are provided in the Supplementary
Methods.

\subsection*{Model benchmarking}

To assess the performance of the proposed cardinality-optimised patient
recruitment and sample allocation framework, we conducted a structured
benchmarking analysis against non-optimised baseline strategies. The
evaluation was designed to separately examine recruitment performance
and sample allocation performance while maintaining consistency in
cohort definitions and input data. Full pseudocode for all baseline
algorithms is provided in Supplementary Section 'Baseline Algorithms
for Recruitment and Sample Allocation'.

Given the absence of an established integrated comparator for this
combined planning problem, the proportional and greedy baselines were
selected as realistic staged planning alternatives, representing both
practical rule-based approaches and stronger optimisation-inspired
greedy heuristics commonly used for marginal-gain and set-cover allocation
problems. Recruitment performance was evaluated by comparing the cardinality-optimised
model against two alternative recruitment strategies: a proportional
heuristic recruitment baseline (Algorithm S1) \citep{chvatal_greedy_1979}
and a greedy recruitment baseline (Algorithm S2) \citep{nemhauser_analysis_1978}.
The objective was to compare the total number of patients recruited
under each method as the number of participating healthcare providers
(HCPs) increased from single-centre inclusion to full multi-centre
participation.

In the proportional heuristic recruitment strategy (Algorithm S1),
patients were assigned to hospitals in proportion to site-specific
patient availability. For each disease, weights proportional to the
estimated patient counts at the selected hospitals were computed and
used to distribute the disease-specific recruitment target, subject
to hospital-level capacity constraints and the predefined lower and
upper bounds on disease cohort sizes. If the initial proportional
allocation did not satisfy the recruitment target for a disease, a
rebalancing loop iteratively assigned remaining patients to hospitals
with residual capacity and sufficient available patient counts. This
strategy prioritises feasibility and local availability but does not
incorporate global coordination across diseases or sparsity considerations.

In addition, a greedy recruitment baseline (Algorithm S2) was implemented.
This approach iteratively assigns patients to disease--hospital pairs
that maximise marginal recruitment gain while respecting hospital
capacity constraints and patient availability limits. At each iteration,
the feasible assignment yielding the largest immediate increase in
recruited patients is selected. Although more structured than the
proportional heuristic, this strategy remains locally greedy and does
not optimise recruitment decisions jointly across diseases.

To evaluate sample allocation performance independently of recruitment
differences, we first solved the integrated recruitment--allocation
optimisation model under site-specific patient availability and laboratory
capacity constraints. To establish a baseline comparison without the
confounding effect of throughput limitations, the model was also solved
under unconstrained laboratory capacity conditions. The resulting
planned disease-specific patient totals, denoted by $y$ were then
fixed and supplied to the allocation baseline algorithms to ensure
that all allocation methods operated on an identical cohort.

In the allocation benchmarking phase, each required biomarker--disease
measurement was treated as a demand for two analytical samples, corresponding
to plasma and urine. Two allocation baselines were considered. The
first, a heuristic disease-wise allocation strategy (Algorithm S3),
applied a set-cover procedure independently for each disease to select
analytical platforms that satisfied biomarker coverage requirements
while limiting the number of platforms used. This approach treats
diseases separately and does not exploit potential cross-disease efficiencies.

The second allocation baseline implemented a global greedy marginal
set-cover strategy (Algorithm S4), following the classical greedy
approximation framework for set covering. In this procedure, analytical
platforms were selected sequentially based on maximising marginal
biomarker coverage per unit cost across all diseases. This global
selection phase was followed by a disease-specific allocation step
restricted to the platforms chosen in the global phase. Although this
method incorporates cross-disease information and cost-awareness,
it remains greedy and does not guarantee global optimality.

Across all allocation experiments, identical biomarker requirements,
cost parameters, and platform capability matrices were used. No re-optimisation
of recruitment was performed in the baseline allocation cases, ensuring
that performance differences reflected allocation strategy rather
than cohort variation.

To assess robustness under operational constraints, progressively
tighter sample throughput limits were imposed on both global and targeted
platforms. Platform capacities were varied from 0 to 2000 samples
per site in increments of 200, and feasibility was evaluated for the
cardinality-optimised model as well as for the heuristic and greedy
allocation baselines.

In addition to benchmarking performance and feasibility, we examined
the model\textquoteright s ability to manage trade-offs between competing
objectives. Specifically, we analysed the interaction between alignment
with hospital-specific clinical interests and minimisation of non-local
sample shipment. A parametric sweep of the clinical interest weight
$\lambda$ was conducted over the range from 0 to 10,000. For each
value of $\lambda$ recruitment patterns and associated logistics
penalties were recorded to characterise the resulting Pareto frontier. 

Together, the HCP engagement scenarios, laboratory capacity sweep,
and clinical-interest weight sweep provide sensitivity analyses for
recruitment scale, operational capacity, and objective-weight calibration.

Detailed pseudocode for Algorithms S1--S4 is provided in Supplementary
Section 'Baseline Algorithms for Recruitment and Sample Allocation'.

\subsection*{Ethics approval and informed consent}

The hospital-provided estimates of patient counts used in this manuscript
to assess the feasibility of clinical study recruitment and downstream
research analysis contained no personal or patient-identifiable information
and could not be linked to individuals. The data were fully anonymous
and comprised only aggregate estimates of recorded cases for each
disease or the numbers potentially available for recruitment at each
hospital. Consequently, ethics approval and consent were not required
for this recruitment-feasibility assessment, as it involved no human
participants, no interventions, and no identifiable data.

The analyses reported here relate solely to recruitment-feasibility
modelling and do not form part of the Recon4IMD Clinical Study. Within
the Recon4IMD project, the terms governing collaboration and data
transfer between all members, including collaborating hospitals, have
been formally established through legal agreements. Under this framework,
Recon4IMD members act as joint data controllers and processors.

The Recon4IMD Clinical Study is a separate prospective investigation
that will involve patient recruitment, informed consent, biological
sample collection, and the transfer of pseudonymised clinical data.
Ethical approval for the clinical study is independently governed
and has already been granted at several sites. At T\"ubingen approval
was granted by the Ethics Committee at the Faculty of Medicine of
the Eberhard Karls University and University Hospital T\"ubingen
with Approval Number 371/2025BO2. At Heidelberg approval was granted
by the Ethics Committee of the Medical Faculty, Heidelberg University
with Approval Number S-645/2024. In the United Kingdom approval was
granted by the West Midlands Coventry and Warwickshire Research Ethics
Committee with Approval Number 25/WM/0032. Recruitment at any site
will only begin once the required local ethical approval has been
obtained.

\section*{Data Availability}

The clinical recruitment and metabolomic lab capability datasets analyzed
during this study are available in the GitHub repository at \url{https://github.com/Digital-Metabolic-Twin-Centre/clinical-study-optimisation}.
To protect participant privacy and institutional profiles, certain
sensitive operational data, specifically the raw estimated patient
counts per healthcare provider and the specific allocation of samples
to individual laboratories, have been withheld in their raw form.
These data were processed using differential privacy ($\varepsilon=1.5$)
to ensure anonymity while preserving analytical utility. Reference
data for inherited metabolic diseases, causal genes, and metabolite
biomarkers were sourced from public databases: the Virtual Metabolic
Human (VMH) \citep{noronha_virtual_2019}, IEMbase \citep{ferreira_proposed_2019},
and the International Classification of Inherited Metabolic Disorders
(ICIMD) \citep{ferreira_international_2021}. Any additional data
supporting the findings of this study are available from the corresponding
author upon reasonable request.

\section*{Code Availability}

The complete implementation of the optimisation framework presented
in this study is publicly available at \url{https://github.com/Digital-Metabolic-Twin-Centre/clinical-study-optimisation}.
This repository includes a MATLAB Live Script (compatible with R2024a)
providing end-to-end execution of the model, along with the anonymised
datasets. The framework utilizes a cardinality optimisation function
\citep{fleming_cardinality_2023a}, available through the COBRA Toolbox
\citep{heirendt_creation_2019c} and requires the GUROBI solver \citep{gurobioptimizationllc_gurobi_2024}.
Detailed documentation of mathematical formulations and testing routines
is provided in the repository to ensure full reproducibility.

\section*{Acknowledgements}

This study was funded by the European Union\textquoteright s Horizon
Europe Framework Programme (grant agreement No 101080997), the Swiss
State Secretariat for Education, Research and Innovation (grant No
23.00232), United Kingdom Research and Innovation (grant Nos 10083717
and 10080153), and the Deutsche Forschungsgemeinschaft (DFG, German
Research Foundation) under Germany's Excellence Strategy (CIBSS -
EXC-2189 - Project ID 390939984) to L.H, and Czech grant projects
(BBMRI\_CZ LM2023033, RVO VFN64165).

\section*{Author Contributions}

F.Z. conceived and designed the study, developed the mathematical
framework, cleaned and analysed data, and drafted the manuscript.
C.C. communicated with healthcare providers and compiled the data.
T.D. helped with the IMD list and reviewed the manuscript. M.M. provided
patient numbers, reviewed the IMD list, and provided clinical insight.
J.T. helped with the IMD list and with the definition of the Diagnostic
Confidence Score. J.M. analysed and compared the metabolomics data
and communicated with the metabolomic labs.

P.J., S.W.G., U.S., J.J.O\textquoteright B., T.T., M.L.C., J.d.l.H.,
N.D., D.M., P.W., I.K., G.L.M., A.T.v.d.P., H.H.H., T.G.J.D., F.J.v.S.,
S.D.K.K., A.M.D., C.D.V., C.D.v.K., J.B.H., S.O\textquoteright S.,
S.A.F., and M.S. contributed IMD data.

L.H., M.R.H.F., M.J.N., I.M.L.W.K.K., K.B.P.E., G.J.G.R., J.A.C.,
J.J.M.J., A.H., and T.H. provided metabolomic platform capabilities.
S.R. contributed patient numbers and reviewed the manuscript. S.K.
contributed IMD data, reviewed the manuscript, and provided patient
participation statistics. I.T. provided the VMH data and contributed
to the manuscript review. R.M.T.F. supervised the research, contributed
to the development of the mathematical model, provided guidance on
study design and methodology, contributed to the interpretation of
results, and revised the manuscript.
\section*{Competing Interests}

The authors declare no conflicts of interest.\bibliographystyle{naturemag}
\bibliography{references}

\end{document}